\documentclass[manuscript,screen,nonacm]{acmart}

\usepackage[utf8]{inputenc}
\usepackage{url}
\usepackage{wrapfig} 
\usepackage{graphicx}
\usepackage{pdfpages}
\usepackage{multirow}

\AtBeginDocument{%
  \providecommand\BibTeX{{%
    \normalfont B\kern-0.5em{\scshape i\kern-0.25em b}\kern-0.8em\TeX}}}

\begin{document}

\title{User Study: Comparison of Picture Passwords and Current Login Approaches}

\author{Ignacio Astaburuaga}
\email{Ignaciochg@nevada.unr.edu}
\orcid{0000-0001-6635-9946}
\affiliation{%
  \institution{University of Nevada, Reno}
  \city{Reno}
  \state{Nevada}
  \country{USA}
}


\renewcommand{\shortauthors}{Astaburuaga}

\begin{abstract}
In this research, we conduct a user study that compares different computer/system authentication methods. More specifically, we look into comparing regular password authentication with picture authentication. Picture authentication means selecting a sequence of pictures from a set of pictures (30). We present users with both interfaces; various metrics are tracked while the participants conduct a variety of user authentication-related tasks. Other metrics include user perception of security with such technologies.
\end{abstract}

\ccsdesc[500]{Human-centered computing~User studies}
\ccsdesc[300]{Human-centered computing~Empirical studies in HCI}
\ccsdesc[500]{Security and privacy~Systems security}


\maketitle

\section{Introduction}

Cybersecurity of computer systems is a very prominent field. Today computers are intertwined with everyone's life and can access sensitive information. Most users tend to treat computers as if they don't play such a significant role in their lives. In the event of a compromise, they can have devastating results. For this reason, we look into ways to improve security in a way that is not invasive to users but is sufficient and of high quality to protect users. Currently, multiple authentication factors are used, including username and password, PIN, SMS, app, face recognition, and fingerprint. Generally, we look into analyzing how each authentication method performs under each task. We hope to use this data to further the research in the Multi-Factor Authentication (MFA) scenario. Multi-factor authenticaton (MFA) or two-factor authentication (2FA) is a way for the user to provide and authenticate its identity using multi-factor of  authentication. This is especially important as, in the current state of security of systems, hackers find breaking into systems increasingly easy as users use weak one-factor authentication. Combining multiple authentication factors will drastically reduce the chances of unauthorized access by malicious actors.

Multi-factor authenticaton (MFA) or two-factor authentication (2FA) is a way for the user to provide and authenticate its identity using multi-factor of authentication. This is especially important as, in the current state of security of systems, hackers find breaking into systems increasingly easy as users use weak one-factor authentication. Combining multiple authentication factors will drastically reduce the chances of unauthorized access by malicious actors.

 We conducted this research in multi-factor authentication components with HCI in mind as another mechanism to prevent cyber attacks by better understanding how we use these technologies.
For this reason, we also look into biometrics, which is particularly interesting in cybersecurity. It is one of the tools we have to combat some of the shortcomings of technology.

Biometrics are body measurements, calculations, and statistical analysis related to human characteristics. Generally speaking, this is done by measuring people's physical and behavioral traits \cite{chow_def, wiki_def}. The field of biometrics came about purely out of necessity, that is, to correctly identify individuals repeatedly.  The grand majority of HCI and authentication is related to human behavior. Mostly done using biometrics, for example, using behavior for authentication purposes \cite{and1_gesture}, achieved via VR technology \cite{vr1,vr2,vr3_motion}, touchscreen utilization \cite{and1_gesture, and2, and3}, user movements like eye tracking movements \cite{vr3_motion}.  Cybersecurity for users is a tradeoff between usability and security, where for most users it is all about the former. Although reliable and accurate, biometrics is very cumbersome to users, making it a bad candidate for multi-factor authentication.

In this research, we conduct a user study that compares different computer/system authentication methods. More specifically, we look into comparing regular password authentication with picture authentication (as detailed in NIST's NISTIR 7030 documentation standard \cite{pic-pass}). Picture authentication means selecting a sequence of pictures from a set of pictures (30). We present users with both interfaces; various metrics are tracked while the participants conduct a variety of user authentication-related tasks. Other metrics include user perception of security with such technologies.

\section{Methodology}
\subsection{Independent variable and factors}
We have compared two authentication methods: regular password and picture password. We all use the regular passwords daily to authenticate to our services. Picture password is a similar concept where we define the password in terms of pictures on the screen instead of characters on a keyboard. We select a sequence of pictures from a set of thirty pictures. 

The independent factor of this study is the "authentication method." In this case, the two authentication factors are "regular password" and "picture password ." Picture passwords can be either randomized picture order or non-randomized, and they can also have a theme or be single picture mode. This gives us a total of four possible picture password configurations for five possible interfaces. We have decided to test three authentication methods: Regular Password, Picture Password with an animal theme, and Picture Password with a single image. Hereafter, these methods may be referenced as classic, them, and single, respectively.

\subsection{Participants}
As our research focuses on finding the best-suited authentication factors for the common technology user, we conducted our study with various individuals representing a technology-using population. We will not focus on only including tech-savvy or individuals in cybersecurity fields. This, however, does not mean that our group of participants did not include such individuals. We aim to capture realistic measurements for a regular technology consumer and capture a diverse selection of male and female college-level educated individuals. Individuals were recruited through word of mouth at the University of Nevada, Reno campus.

We conducted this study with 12 participants; we discuss how these are divided in a future section.

\subsection{Research questions/Hypothesis}
We propose the following questions to be answered by this study:

H1. Picture passwords reduce password fatigue/stress. 

H1.2 Users prefer to use picture passwords to picture passwords for simplicity.

H1.2.1 Users have an increased perception of increased security by using picture passwords, compared to regular passwords.

H1.3 Picture passwords with a single picture theme can reduce password fatigue.

H1.4 The users prefer Theme Picture Password theme mode interface.

H1.5 The users prefer Single Picture Password theme mode interface.

H2. Users can log in faster using picture passwords than regular passwords. We will measure account creation, login time, and error rate(described later). 

H3. Users can retain more picture passwords than regular passwords. Test users on creating multiple accounts and see if adding one more account will make them lose retention of the previously learned account, and find that breaking point. 

H4. Password reuse is reduced in picture passwords. We will test it by having them come up with multiple passwords in short succession and measure the similarity between passwords. How much do these passwords vary from the set? We will use a similarity score for their created set. 

H5. Password sharing is significantly reduced with picture passwords. We share a password with the user and let them type it in, measuring their error rates.

H6.  Picture passwords for Account creation are preferred.

H7. Users have a more straightforward use with picture passwords (frustration due to error); in other words, it reduces errors. We will check for error rates over tries. (error-rate/total-submits)

H8. On average more users remembered passwords within 20 minutes of making (after being asked to do other tasks) them with picture passwords (T2a,b).

H9. The amount of user interaction in the picture password interface is reduced (interaction score = clicks+keys). This could lead to the reduction of the perception of load.

H10. The number of mistakes (user identified) is reduced for picture passwords (number of passwords cleared or backspace used).

H11. The complexity of passwords is inherently higher with picture passwords with the only requirement being eight in length. Rank password complexity into the score and look into password shortcuts or patterns that make it inherently insecure.

\subsection{Tasks}
In the task descriptions (below), if it describes a password, picture, or both, it means that the task will be repeated using a password, picture password, or both. This does not mean we will mix both passwords in the same task; it means users will have to execute the same task using different password interfaces. Each task was designed to correspond to one or more of the above hypotheses for testing purposes. 

We will measure metrics for all account creation and login tasks. These metrics include time to successful completion time, number of times the participant used the system incorrectly, which will be measured in backspace usage, password clear button usage, and incorrect number of attempts. Other metrics include the number of buttons and clicks pressed. We will also use an algorithm to score the password sets entered by the user to check for similarity within the set.

\begin{enumerate}

    \item When participant agrees to the study, they will sign up for a time slot with the study coordinator.
    \item Two days before the study, an instructional document will be sent to participant, which they can review at their leisure. 
    \item The participant meets up with the study coordinator at the research lab at the agreed time slot.
    \item They will receive a pre-study questionnaire that they will be requested to fill out (virtual document).
    \item Once complete, they will receive instructions on the study from the study coordinator on how to perform the tasks. The coordinator will also demonstrate how to use the interfaces.
    \item The participant will be asked if they would like to practice using the Picture Password interface. If they do, they will be given access to the demonstration interface previously shown (5 minutes maximum).
    \item Once the instructions are provided, the study begins.
    \item The participant will be provided an account creation interface. The interface will be different given the independent variable being tested.
    \item The participant must quickly create the account. (This account is not tied to anything and user will be told explicitly not to use any of his current account information for this study.)
    \item After account creation, the user will be instructed to log in using one of the interfaces being tested.
    \item Once they have successfully done so, they must perform the experiment again with different parameters.
    \item The study software will navigate the participants between each of the configurations and tasks during this sequence.
    \item Once the participants are done with all the tests above, the user study has concluded.
    \item The participants are then asked to provide feedback in a post-study questionnaire (virtual document). During this step, participants can provide quantitative and qualitative feedback on their experiment experience alongside suggestions for improvement.
    \item Once the post-study questionnaire is completed, the participant's job is complete, and they may conclude their participation.
\end{enumerate}
Here is a more extensive list of tasks that correspond to steps 7-10:

T1. Users will be asked to create an account (using either password, picture, or both). This will be an ongoing task that will test the user's ability to recall passwords after some time or other tasks; Users will be prompted to remember them. Users will be asked to enter the password at the end of all other tasks. We will measure time, incorrect usage, clicks, button presses, backspaces, clears, and errors. 

T2. Users will be asked to create a single password (either password, picture, or both). The user will then be asked to write it down as he was going to share it with someone (certain rules are set here). Next, the user will be asked to forget about the password. Towards the end of the session with the user, he will be asked to read his or her notes and log in. The same metrics will be measured as all other tasks. 

T4. The participant will be asked to create an account (password, picture, or both), and metrics will be taken on this account creation task. The user will be asked to log in. The participant will be asked to log in five times using the same credentials (for each method). The same metrics will be measured as all other tasks. In this task, we test the account creation speed and log-in speeds more specifically.

T5. If the user is using picture password for T4. The picture password location will be randomized to test added security. The participant will be asked to log in another 5 times. This time the picture locations will be randomized and not the same as when they created them or previously logged in. 

T6. Users will be asked to create an account and log in using specific password hardness/security requirements. We will then test the highest level of security so users can still successfully log in by increasingly asking them to increase their password security. The highest level of security will be a metric we will measure, as well as all other tasks.

T6. The participant will be asked to create a password (either password, picture, or both). We will then test for retention when multiple passwords are being memorized. We ask the participant to log in. We will then ask the user to create one more account and ask them to log in to all of the previously created accounts on this task. We will repeat the cycle of adding one more account until the user can no longer successfully log in to the accounts. The same metrics will be measured as all other tasks. In this case, the number of successfully retained accounts will also be another metric we will use.

T7. Predefined accounts will be created before user sessions are held. These will include a range of regular and picture passwords. Participants will be given a post-it or piece of paper with the password, which will simulate someone handing them the password. Participants will be asked to log in using that information. The same metrics will be measured as for all other tasks.

The entry and exit questionnaires are located in the appendix of this paper. 

\subsection{Apparatus}

\begin{figure}[h]
\caption{Study computer workstation.}
\centering
\includegraphics[angle=-90, width=0.35\textwidth]{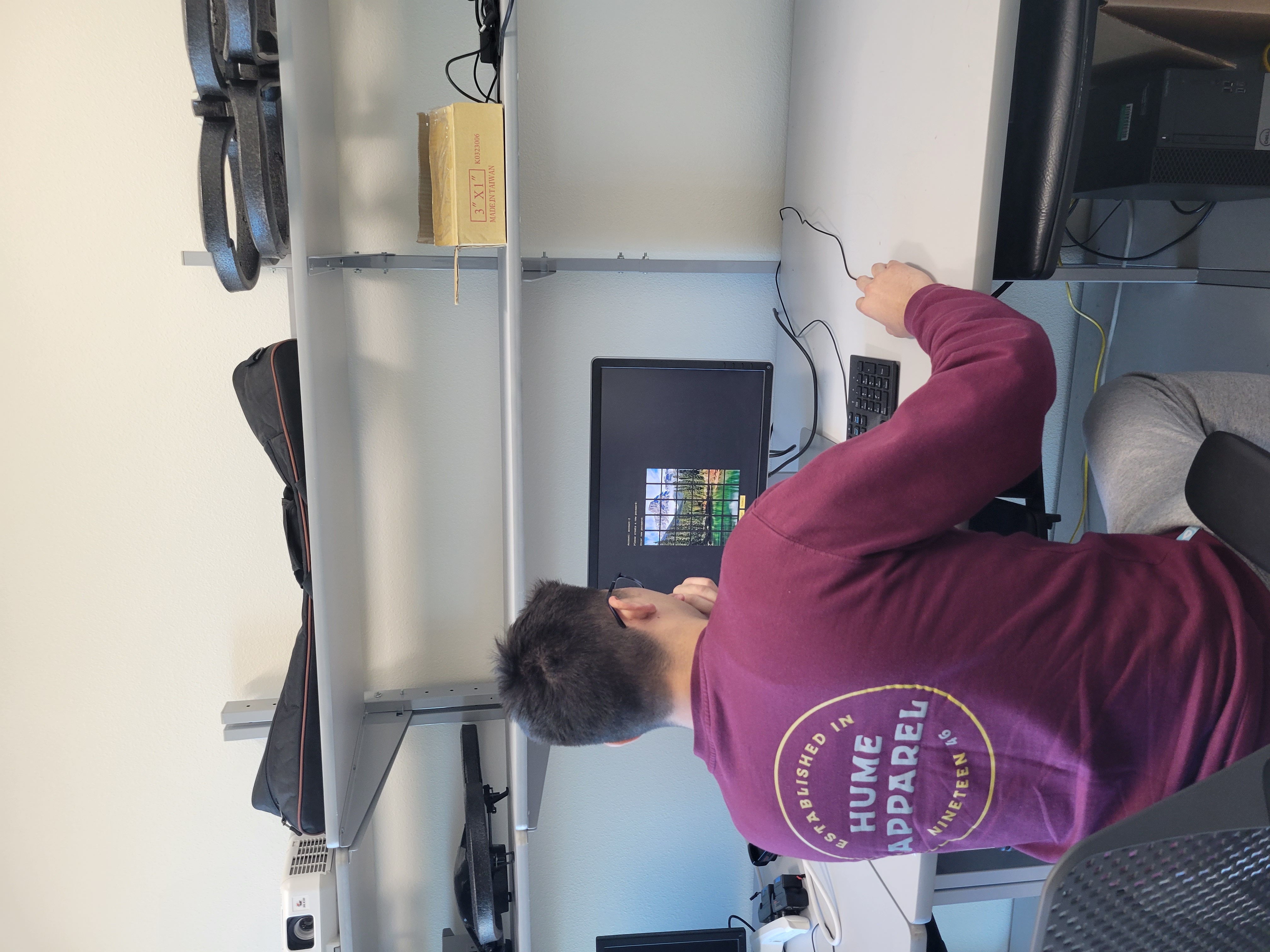}
\label{fig:setup}
\end{figure}

\begin{figure}[h!]
\caption{Example of regular password interface using.}
\centering
\includegraphics[width=0.4\textwidth]{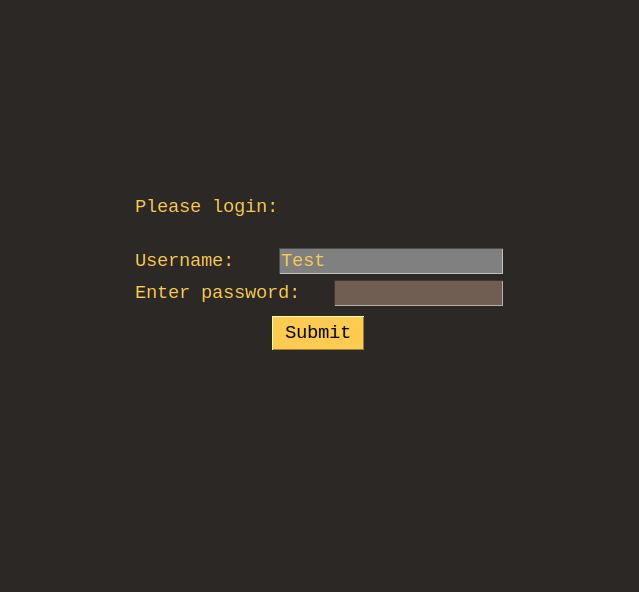}
\label{fig:classic}
\end{figure}

\begin{figure}[h!]
\caption{Example of picture password interface using dog and cats theme.}
\centering
\includegraphics[ width=0.4\textwidth]{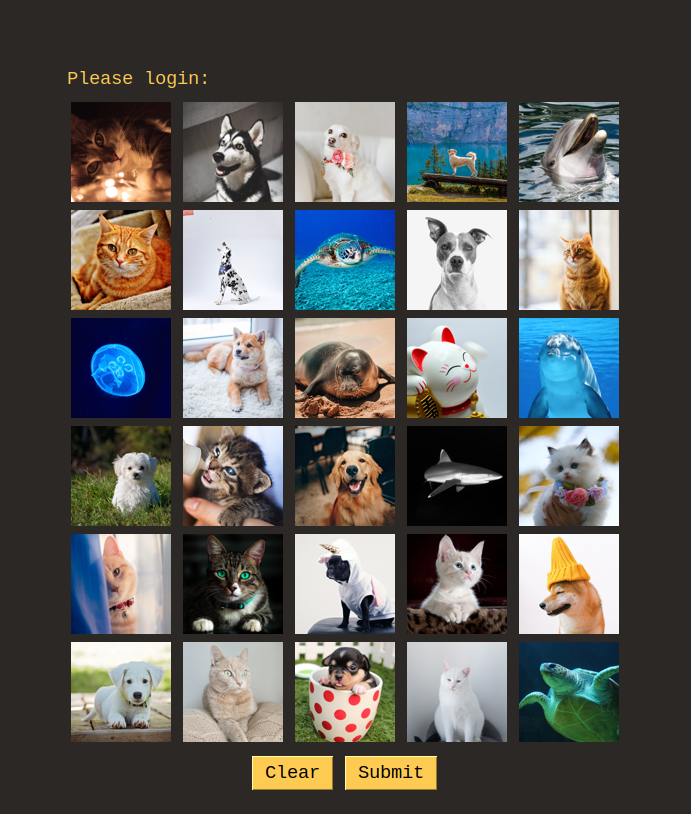}
\label{fig:pic-pass-theme}
\end{figure}

\begin{figure}[h!]
\caption{Example of picture password interface using sea shore as single picture mode.}
\centering
\includegraphics[width=0.4\textwidth]{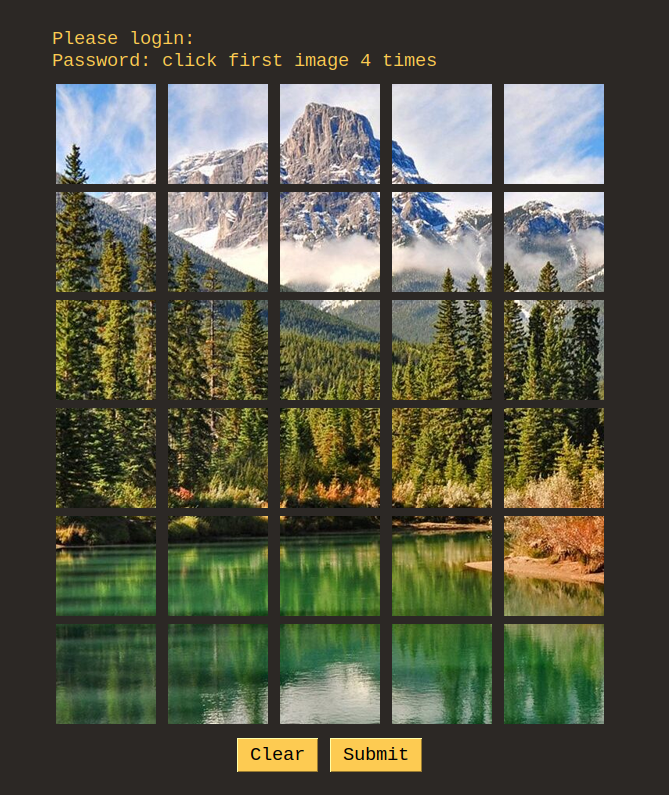}
\label{fig:pic-pass-single}
\end{figure}

Users were asked to conduct account creation and login tasks as described in the previous section. This user study was conducted with IRB approval. We used a Dell Optiplex 7070 computer to conduct the user study. We installed Linux operating system. We also implemented a custom software interface that implements account creation, login and metric tracking for both regular passwords and picture passwords. The picture password interface implemented is described in the NIST NISTIR 7030 documentation standard \cite{pic-pass}. The setup can be seen in figure \ref{fig:setup}, and the interface samples can be seen in figures \ref{fig:classic},\ref{fig:pic-pass-theme} and \ref{fig:pic-pass-single}. 

The user study was conducted in a quiet room side room of a campus research lab. This place was chosen because we have access to it exclusively. Therefore, we can conduct all the studies in a controlled and undisturbed manner. 

\subsection{Procedure}
Before the participants were convened to go through the study, we had two participants go through the study to find any possible improvement to the study workflow as a pilot demo. These two individuals were not part of the main study and were purely to test the instructions, software, and surveys before they were used with the rest of the participants. We also used these trial runs to adjust the session lengths, number of interfaces per user, and number of tasks.

\begin{itemize}
\item A participant arrives at the lab; we greet them and guide them to the testing room where the computer is. 
\item We briefed them. In this briefing, we asked the user to conduct the pre-study survey.
\item After a participant finished the survey, we let them know what we needed them to do in this study.
\item We instructed them to create accounts and login in a repeated way, and we need them to do it in a precise and concise manner.
\item The examiner asked the participant if he or she had any questions and answered them in a controlled manner. 
\item We started by showing participants a picture password; we demonstrated what the picture password is and how it works.
\item We let the user do 5 minutes of practice with the picture password interface (all participants opted out).
\item We started the tasks as described in the task section. The user was asked to create accounts and login into the accounts using three different login methods. The software was self-guided with instructions. 
\item After the participant is finished, the software prompted them to seek the examiner. The examiner handed the participant \item the post-study survey, which they completed. 
\item The examiner asked the participants if he or she had any questions. 
\item The participants were thanked for their time and were dismissed. 
\end{itemize}

\subsection{Design}

We have gathered the following metrics (dependent variables) for all login and account creation tasks:

\begin{itemize}
    \item Total time (per action and task)
    \item Error
    \item Password retention
    \item Password shareability
    \item number of clicks
    \item Key-presses 
    \item Number of Enter key presses
    \item Number of Backspace key presses
    \item Number of Clear (erasing password)
    \item Number of skipped logins, tracks giving up
    \item Total number of submissions
    \item Number of incorrect submissions
    \item Number of rounds for Task 6 (we can also use this to tack persistence)
    \item Passwords and usernames used
    
\end{itemize}

We have additionally calculated the $error rate$ as $incorrectlySubmitted/totalSubmitted$.

For non-parametric dependant variables, we collected the following using the questioners:
\begin{itemize}
    \item Stress/Fatigue levels
    \item Perception of security
    \item Preference of interface (in two forms)
    \item Password 
\end{itemize}

The data collected was more than the data we analyzed in the next section. Since this study studies a single independent variable with three levels, we used between subject groups with counterbalancing as shown in Table \ref{tbl:counter}. We chose to use this method because we felt that the Picture Password was already at a disadvantage since Regular Password is used by all users in their normal lives.

For the remainder of this paper, Regular, Picture with theme pictures, and Picture with a single picture will be addressed as classic, theme, and single, respectively.

\begin{table}[]
  \caption{Counterbalancing of groups.}
  \label{tbl:counter}
\centering
\begin{tabular}{cc}
\hline
Group & Order \\ \hline
1 & Classic, Theme, Single \\
2 & Classic, Single, Theme \\
3 & Theme, Classic, Single \\
4 & Theme, Single, Classic \\
5 & Single, Classic, Theme \\
6 & Single, Theme, Classic
\end{tabular}
\end{table}

\section{Results \& Analysis}
\subsection{Research Questions}
We have collected data to analyze regarding all of the hypothesis aforementioned, but due to time constraints, we will only look at results originating from data regarding the following hypothesis:

H1. Picture passwords reduce password fatigue/stress. 

H1.2.1 Users have an increased perception of increased security by using picture passwords, compared to regular passwords.

H1.4 The users prefer Theme Picture Password theme mode interface.

H1.5 The users prefer Single Picture Password theme mode interface.

H2. Users can log in faster using picture passwords than regular passwords.

H7. Users have a more straightforward use with picture passwords (frustration due to error); in other words, it reduces errors. We will check for error rates over tries. (error-rate/total-submits)

\subsection{Data}
In this analysis section, we only look at account creation time (average across tasks), logging time (Task 3), error rate across all tasks, and the results from the questionnaire.

\begin{figure}[h!]
\centering
\includegraphics[trim={0 2cm 0 2cm},clip,width=\textwidth]{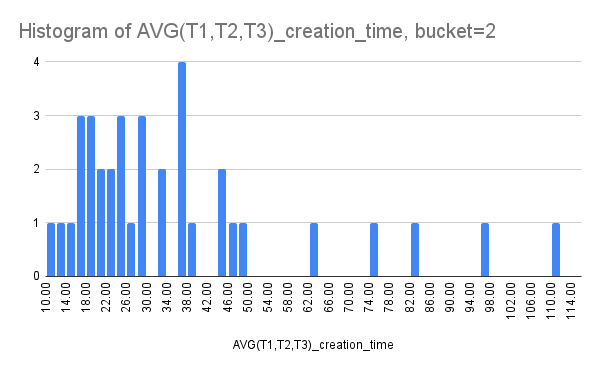}
\caption{Histogram of account creation time data, in seconds. Bucket size of 2s.}\label{fig:hist-create}
\end{figure}

\begin{figure}[h!]
\centering
\includegraphics[trim={1.2cm 0.6cm 0 2cm},clip,width=\textwidth]{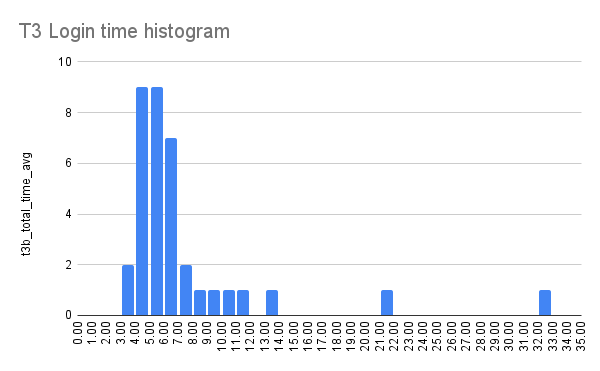}
\caption{Histogram of log-in time data in seconds.}
\label{fig:hist-login}
\end{figure}

\begin{figure}[h!]
\centering
\includegraphics[trim={0 1.5cm 0 2cm},clip,width=\textwidth]{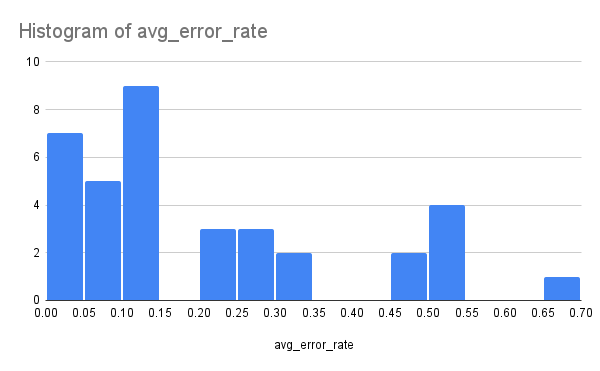}
\caption{Histogram of error rate data, in percentage.}
\label{fig:hist-error}
\end{figure}

The results are shown in Tables \ref{tbl:creation_time}, \ref{tbl:login_time},\ref{tbl:error_data}, \ref{tbl:survey-data} and their corresponding histograms in Fig. \ref{fig:hist-create}, \ref{fig:hist-login}, \ref{fig:hist-error}. By observing each group's means by data type, we can see that counterbalancing did not work (Discussed on the Discussion Section).

\begin{table}[]
  \caption{Average(T1,T2,T3) Creation Time.}
  \label{tbl:creation_time}
\centering
\begin{tabular}{ccccccc}
\hline
Participant & Classic & Theme & Single & Group & Mean & SD \\ \hline
1 & 15.79 & 17.32 & 12.78 & \multirow{2}{*}{1} & \multirow{2}{*}{40.33} & \multirow{2}{*}{33.33} \\
7 & 63.64 & 96.05 & 36.39 &  &  &  \\ \cline{5-7} 
2 & 11.34 & 32.93 & 16.72 & \multirow{2}{*}{2} & \multirow{2}{*}{29.78} & \multirow{2}{*}{15.20} \\
8 & 49.17 & 44.74 & 23.77 &  &  &  \\ \cline{5-7} 
3 & 29.12 & 83.95 & 111.75 & \multirow{2}{*}{3} & \multirow{2}{*}{51.58} & \multirow{2}{*}{37.28} \\
9 & 22.26 & 24.56 & 37.82 &  &  &  \\ \cline{5-7} 
4 & 29.02 & 28.53 & 21.34 & \multirow{2}{*}{4} & \multirow{2}{*}{26.94} & \multirow{2}{*}{6.67} \\
10 & 27.14 & 37.37 & 18.23 &  &  &  \\ \cline{5-7} 
5 & 33.36 & 74.82 & 44.79 & \multirow{2}{*}{5} & \multirow{2}{*}{35.55} & \multirow{2}{*}{21.81} \\
11 & 17.55 & 18.02 & 24.73 &  &  &  \\ \cline{5-7} 
6 & 39.312 & 18.51 & 20.09 & \multirow{2}{*}{6} & \multirow{2}{*}{30.77} & \multirow{2}{*}{11.46} \\
12 & 24.00 & 36.40 & 46.32 &  &  &  \\ \hline
Mean & 30.14 & 42.77 & 34.56 &  &  &  \\
SD & 14.79 & 27.16 & 26.75 &  &  & 
\end{tabular}
\end{table}

\begin{table}[]
  \caption{Average log-in time across three repetitions of Task 3.}
  \label{tbl:login_time}
\centering
\begin{tabular}{ccccccc}
\hline
Participant & Classic & Theme & Single & Group & Mean & SD \\ \hline
1 & 3.03 & 4.75 & 3.86 & \multirow{2}{*}{1} & \multirow{2}{*}{4.04} & \multirow{2}{*}{1.58} \\
7 & 1.78 & 6.45 & 4.36 &  &  &  \\ \cline{5-7} 
2 & 4.43 & 8.19 & 7.70 & \multirow{2}{*}{2} & \multirow{2}{*}{5.63} & \multirow{2}{*}{1.90} \\
8 & 5.47 & 4.57 & 3.44 &  &  &  \\ \cline{5-7} 
3 & 6.36 & 8.13 & 20.50 & \multirow{2}{*}{3} & \multirow{2}{*}{9.21} & \multirow{2}{*}{6.07} \\
9 & 11.14 & 5.14 & 3.97 &  &  &  \\ \cline{5-7} 
4 & 9.42 & 6.83 & 6.85 & \multirow{2}{*}{4} & \multirow{2}{*}{5.62} & \multirow{2}{*}{2.64} \\
10 & 4.07 & 1.84 & 4.68 &  &  &  \\ \cline{5-7} 
5 & 4.85 & 4.35 & 4.89 & \multirow{2}{*}{5} & \multirow{2}{*}{4.97} & \multirow{2}{*}{0.79} \\
11 & 3.99 & 6.175 & 5.60 &  &  &  \\ \cline{5-7} 
6 & 13.56 & 3.47 & 4.51 & \multirow{2}{*}{6} & \multirow{2}{*}{7.08} & \multirow{2}{*}{3.57} \\
12 & 6.43 & 8.19 & 6.31 &  &  &  \\ \hline
Mean & 6.21 & 5.67 & 6.39 &  &  &  \\
SD & 3.48 & 2.01 & 4.62 &  &  & 
\end{tabular}
\end{table}

\begin{table}[]
  \caption{Average(T1,T2,T3) Error Rate Percentage.}
  \label{tbl:error_data}
\centering
\begin{tabular}{ccccccc}
\hline
Participant & Classic & Theme & Single & Group & Mean & SD \\ \hline
1 & 0.31 & 0.055 & 0.055 & \multirow{2}{*}{1} & \multirow{2}{*}{0.19} & \multirow{2}{*}{0.18} \\
7 & 0.52 & 0.11 & 0.10 &  &  &  \\ \cline{5-7} 
2 & 0.00 & 0.10 & 0.69 & \multirow{2}{*}{2} & \multirow{2}{*}{0.29} & \multirow{2}{*}{0.25} \\
8 & 0.47 & 0.23 & 0.23 &  &  &  \\ \cline{5-7} 
3 & 0.45 & 0.00 & 0.15 & \multirow{2}{*}{3} & \multirow{2}{*}{0.15} & \multirow{2}{*}{0.15} \\
9 & 0.07 & 0.10 & 0.15 &  &  &  \\ \cline{5-7} 
4 & 0.00 & 0.05 & 0.00 & \multirow{2}{*}{4} & \multirow{2}{*}{0.13} & \multirow{2}{*}{0.20} \\
10 & 0.21 & 0.52 & 0.00 &  &  &  \\ \cline{5-7} 
5 & 0.00 & 0.10 & 0.00 & \multirow{2}{*}{5} & \multirow{2}{*}{0.19} & \multirow{2}{*}{0.19} \\
11 & 0.50 & 0.27 & 0.27 &  &  &  \\ \cline{5-7} 
6 & 0.31 & 0.50 & 0.15 & \multirow{2}{*}{6} & \multirow{2}{*}{0.24} & \multirow{2}{*}{0.15} \\
12 & 0.076 & 0.15 & 0.29 &  &  &  \\ \hline
Mean & 0.24 & 0.18 & 0.175 &  &  &  \\
SD & 0.21 & 0.16 & 0.19 &  &  & 
\end{tabular}
\end{table}

\begin{table}
  \caption{Questionnaire data.}
  \label{tbl:survey-data}
\includegraphics[width=\textwidth]{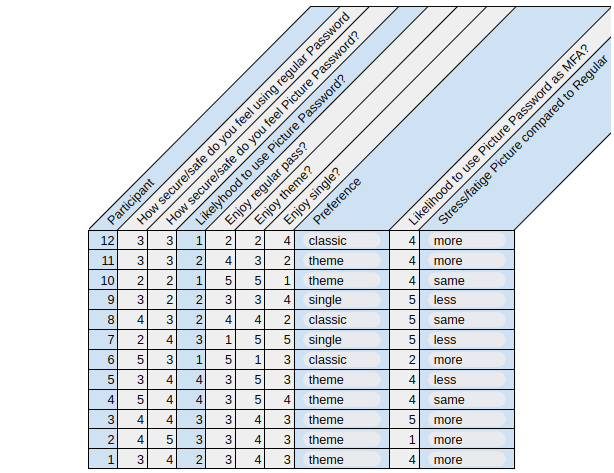}
\end{table}

\subsection{Parametric Tests}
We conducted One-Way ANOVA test on the creation time, log-in time, and error rates. 

The mean account creation time for Classic was 30.14, Theme was 42.77, and Single was 46.32. The difference was statistically significant ($F_{2,33}=0.88, p < 0.5$).

By further investigating the data, as seen in Tables \ref{fig:create-1}, \ref{tbl:create-1}, not all are normally distributed. This distribution has a major effect on the results and tests. We have removed some of the outliers and we can see in Tables \ref{fig:create-2}, \ref{tbl:create-2} that normality improves on both Classic and Single. We can also see that Theme data is heavily right-skewed. The difference was statistically significant ($F_{2,30}=3.13, p \approx 0.05$). We furthered by executing another ANOVA test, as seen on Tables \ref{fig:create-3}, \ref{tbl:create-3} on just Classic and Single. The difference was statistically significant ($F_{1,19}=0.35, ns$). From these tests, we can conclude that we fail to reject $H_0$ and fail to prove H2.

\begin{table}[]
\caption{Creation time data analysis.}
\label{fig:create-1}
\centering
\begin{tabular}{rcccccc}
\cline{2-6}
\multicolumn{1}{c}{} & \multicolumn{5}{c}{Interface} \\ \cline{2-6} 
\multicolumn{1}{c}{} & Classic &  & Theme &  & Single \\ \cline{2-2} \cline{4-4} \cline{6-6} 
\textbf{Skewness} & 1.10 &  & 1.07 &  & 2.47 \\
\textbf{Excess kurtosis} & 1.21 &  & -0.18 &  & 7.05 \\
\textbf{Normality} & 0.30 &  & 0.02 &  & 0.0009 \\
\textbf{Outliers} & 63.64 &  &  &  & 111.75 \\
\textbf{Mean} & 30.14 &  & 42.77 &  & 34.56 \\
\textbf{SD} & 14.79 &  & 27.16 &  & 26.75
\end{tabular}
\end{table}

\begin{table}[]
  \caption{ANOVA results of creation time data.}
  \label{tbl:create-1}
\centering
\begin{tabular}{cccccc}
\hline
Source & DF & Sum of Square & Mean Square & F Statistic & P-value \\ \hline
\textbf{Groups (between groups)} & 2 & 984.94 & 492.47 & 0.88 & 0.42 \\
\textbf{Error (within groups)} & 33 & 18395.43 & 557.43 &  &  \\
\textbf{Total} & 35 & 19380.37 & 553.72 &  & 
\end{tabular}
\end{table}

\begin{table}[]
\caption{Creation time data without outliers analysis.}
\label{fig:create-2}
\centering
\begin{tabular}{rcccccc}
\cline{2-6}
\multicolumn{1}{c}{} & \multicolumn{5}{c}{Interface} \\ \cline{2-6} 
\multicolumn{1}{c}{} & Classic &  & Theme &  & Single \\ \cline{2-2} \cline{4-4} \cline{6-6} 
\textbf{Skewness} & 0.61 &  & 1.07 &  & 0.57 \\
\textbf{Excess kurtosis} & 0.39 &  & -0.18 &  & -1.21 \\
\textbf{Normality} & 0.92 &  & 0.02 &  & 0.17 \\
\textbf{Outliers} & 63.64 &  &  &  & 111.75 \\
\textbf{Mean} & 27.10 &  & 42.77 &  & 27.54 \\
\textbf{SD} & 10.88 &  & 27.16 &  & 11.71
\end{tabular}
\end{table}

\begin{table}[]
  \caption{ANOVA results of creation time data without outliers.}
  \label{tbl:create-2}
\centering
\begin{tabular}{cccccc}
\hline
Source & DF & Sum of Square & Mean Square & F Statistic & P-value \\ \hline
\textbf{Groups (between groups)} & 2 & 2112.6 & 1056.3 & 3.12 & 0.058 \\
\textbf{Error (within groups)} & 30 & 10135.92 & 337.86 &  &  \\
\textbf{Total} & 32 & 12248.53 & 382.76 &  & 
\end{tabular}
\end{table}

\begin{table}[]
\caption{Creation time data of Classic and Single without outliers analysis.}
\label{fig:create-3}
\centering
\begin{tabular}{rccc}
\cline{2-4}
\multicolumn{1}{c}{} & \multicolumn{3}{c}{Interface} \\ \cline{2-4} 
\multicolumn{1}{c}{} & Classic &  & Single \\ \cline{2-2} \cline{4-4} 
\textbf{Skewness} & 0.61 &  & 0.57 \\
\textbf{Excess kurtosis} & 0.39 &  & -1.2 \\
\textbf{Normality} & 0.92 &  & 0.17 \\
\textbf{Outliers} & 63.64 &  & 111.75 \\
\textbf{Mean} & 27.1 &  & 27.54 \\
\textbf{SD} & 10.88 &  & 11.71
\end{tabular}
\end{table}

\begin{table}[]
\caption{ANOVA results of creation time data of Classic and Single without outliers.}
  \label{tbl:create-3}
\centering
\begin{tabular}{cccccc}
\hline
Source & DF & Sum of Square & Mean Square & F Statistic & P-value \\ \hline
\textbf{Groups (between groups)} & 1 & 36.91 & 36.91 & 0.34 & 0.56 \\
\textbf{Error (within groups)} & 19 & 2019.75 & 106.30 &  &  \\
\textbf{Total} & 20 & 2056.67 & 102.83 &  & 
\end{tabular}
\end{table}

As seen in Fig. \ref{tbl:login_time},\ref{tbl:login-time-anova}, Task 3 mean account log-in time for Classic was 6.22, Theme was 5.68, and Single 6.39. The difference was not statistically significant ($F_{2,33}=0.88, ns$). We can conclude that we fail to reject $H_0$ and fail to prove H2.

\begin{table}[]
  \caption{ANOVA results of log-in time data.}
  \label{tbl:login-time-anova}
\centering
\begin{tabular}{cccccc}
\hline
Source & DF & Sum of Square & Mean Square & F Statistic & P-value \\ \hline
\textbf{Groups (between groups)} & 2 & 3.33 & 1.66 & 0.13 & 0.87 \\
\textbf{Error (within groups)} & 33 & 413.29 & 12.52 &  &  \\
\textbf{Total} & 35 & 416.63 & 11.90 &  & 
\end{tabular}
\end{table}

As seen in Tables. \ref{tbl:error_data},\ref{tbl:error-rate-anova}, the data does not follow a normal distribution; therefore, ANOVA is not a suitable test for this data. Meaning we fail to prove H7.

\begin{table}[]
  \caption{Error rate data analysis.}
  \label{tbl:error-rate-anova}
\centering
\begin{tabular}{rcccccc}
\cline{2-6}
\multicolumn{1}{c}{} & \multicolumn{5}{c}{Interface} &  \\ \cline{2-6}
\multicolumn{1}{c}{\textbf{}} & Classic &  & Theme &  & Single &  \\ \cline{2-2} \cline{4-4} \cline{6-6}
\textbf{Skewness} & 0.07 &  & 1.28 &  & 1.87 &  \\
\textbf{Excess kurtosis} & 0.92 &  & 0.17 &  & 4.54 &  \\
\textbf{Normality} & 0.06 &  & 0.01 &  & 0.01 &  \\
\textbf{Outliers} &  &  & 0.52 &  & 0.69 &  \\
\textbf{Mean} & 0.24 &  & 0.18 &  & 0.17 &  \\
\textbf{SD} & 0.21 &  & 0.17 &  & 0.19 & 
\end{tabular}
\end{table}

\subsection{Non-Parametric Test}
In this subsection, we further analyze data from Fig. \ref{tbl:survey-data}. 

We will start by looking into H1.2.1 perception of security. The average for both regular passwords and picture passwords is 3.42 on a scale of 1-5 (1=not safe, 5=most safe). We conducted Wilcoxon signed-ranks test (two-tailed). We conclude that with a $\mu0=0$(no difference) the difference between classic and picture is not big enough to be statistically significant ($Z=-0.073, p=0.94, ns$). Meaning we can not reject the null hypothesis of them being equal. This means that users perceive both regular passwords and picture passwords with equal levels of security; the data also confirms this. Therefore, we reject H1.2.1.

\begin{figure}[h!]
\centering
\includegraphics[trim={0 0.5cm 0 2cm},clip,width=\textwidth]{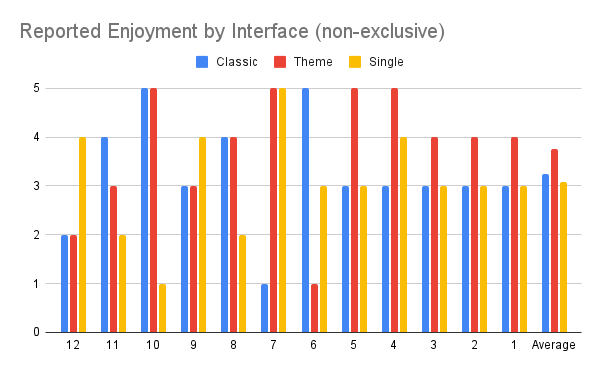}
\caption{Reported Enjoyment by Interface (non-exclusive). Includes Average.}
\label{fig:sec1}
\end{figure}

Next, we will discuss results relating H1.4 and H1.5, preference for picture passwords over regular passwords. Initial observations of the data may lead to believe that users prefer Theme over Single or Classic as seen in Figures \ref{fig:sec1}, \ref{fig:sec1}, these results originate from four different questions. We conducted Friedman Test for Repeated-Measures
and we concluded that it is not statistically significant with $p=.28$ ($\chi^2_r=2.54, ns$). Therefore we fail to reject null hypothesis and can not conclude H1.4 or H1.5.

\begin{figure}[h!]
\centering
\includegraphics[trim={0 1cm 0 2cm},clip,width=\textwidth]{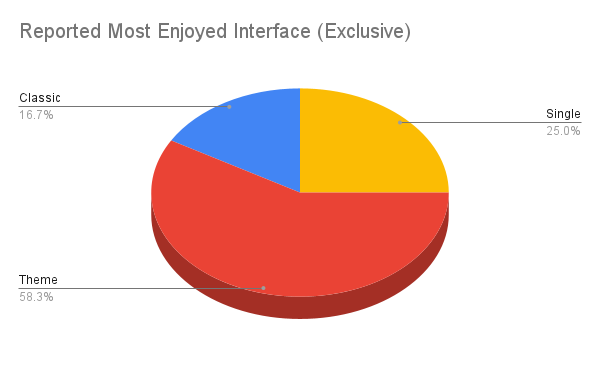}
\caption{Reported Enjoyment by Interface (exclusive).}
\label{fig:sec2}
\end{figure}
\begin{figure}[h!]
\centering
\includegraphics[trim={0 1cm 0 2cm},clip,width=\textwidth]{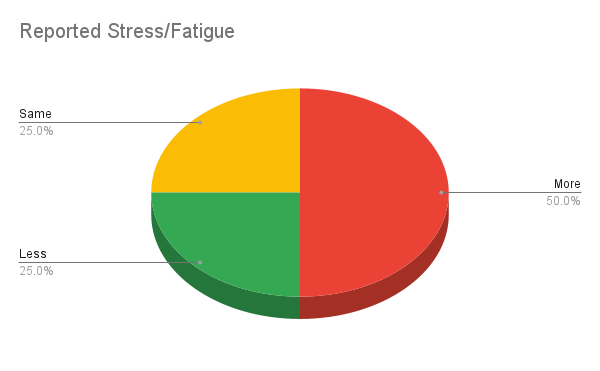}
\caption{Reported Stress/Fatigue pi-chart.}
\label{fig:stress}
\end{figure}
\begin{figure}[h!]
\centering
\includegraphics[trim={0 1.5cm 0 2cm},clip,width=\textwidth]{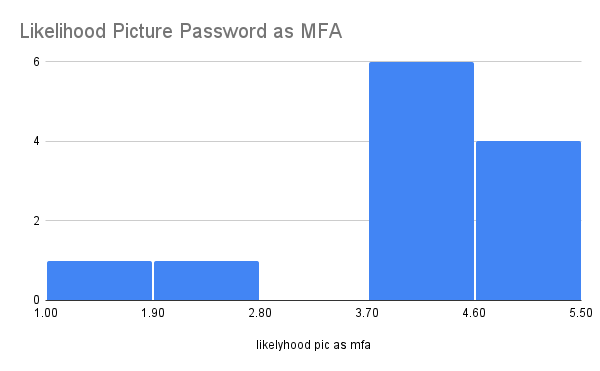}
\caption{Likelihood to use Picture Password as MFA histogram.  1=Not likely, 5=Extremely likely.}
\label{fig:mfa}
\end{figure}
\begin{figure}[h!]
\centering
\includegraphics[trim={0 1.7cm 0 2cm},clip,width=\textwidth]{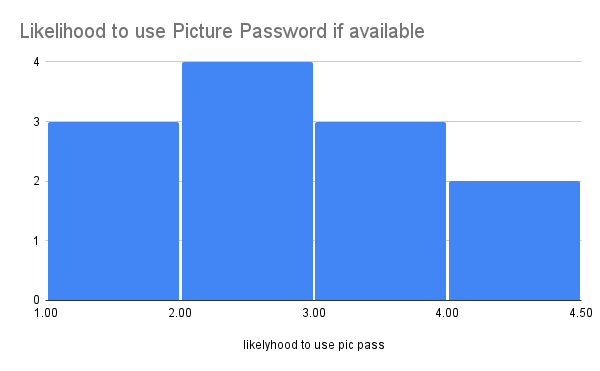}
\caption{Histogram of likelihood to use Picture Password if available histogram. 1=Not likely, 5=Extremely likely.}
\label{fig:available}
\end{figure}


. From the reported qualitative data and figures \ref{fig:stress}, \ref{fig:mfa}, and \ref{fig:available} we can conclude the majority of users in this study would use picture passwords as MFA and would not use them instead of regular passwords. Users reported more password fatigue/stress from picture passwords. We have not conducted significant statistical tests on data for stress/fatigue, likelihood to use picture passwords as MFA, or likelihood to use picture passwords if available. Therefore, we can not say whether the results are true or false (in this case, we can not conclude H1)

\section{Discussion \& Future Work}

One big point here is that outliers made it hard to analyze the data. We can see that most dependent variables in this paper showed either followed a somewhat normal distribution with right skewness or were not normally distributed. For future iterations, it may be wise to define tasks or contain tasks in such a way that eliminates the infinite upper bound. The definition of tasks and execution of the study itself had a huge impact on the data presented in this paper. There are lots of data that were not analyzed, as well as other hypotheses not tested (for which we have the data). 

Since users had no upper bound for time meant that there were outliers, in combination with users not executing the same tasks, in the same manner, every time (perhaps because the tasks were not well constrained) meant that counterbalancing did not work. Therefore resulting in too much variation to make any of the results statistically significant. 

Another important thing to note, we had one user that did not follow instructions and we removed his data from the dataset, therefore we tested 13 individuals. 

Another important note is that picture passwords are at a huge disadvantage compared to regular passwords as regular passwords are used every day by all users. Though the users were granted practice time, it would not be enough to offset their knowledge of regular password usage. Even though we did not conclude with statistically significant results for this comparison, it would be interesting to see how we can overcome this with a longitudinal study.

The ultimate goal of the user study was to compare authentication methods and look at the feasibility of using them in a multi-factor scenario to help improve user security while incurring the least amount of burden. We also looked at the perception of security and fatigue which are contributing factors in users ultimately adopting security measures/technologies. 

In future work, we would like to analyze all the data gathered in this study and perhaps deliberate in some of the other hypotheses mentioned in this paper. We would also like to recreate this study with better-defined tasks in order to reduce variation as stated above. We would like to conduct a study relating Multi-Factor Authentication (MFA) with three factors as one of the user study's independent variables.

\clearpage 
\bibliographystyle{ACM-Reference-Format}
\bibliography{main} 

\newpage
\section*{Appendix}
\appendix
\section{Pre-Study Survey Questionnaire}\label{"apx:pre"}
\includepdf[pages=-]{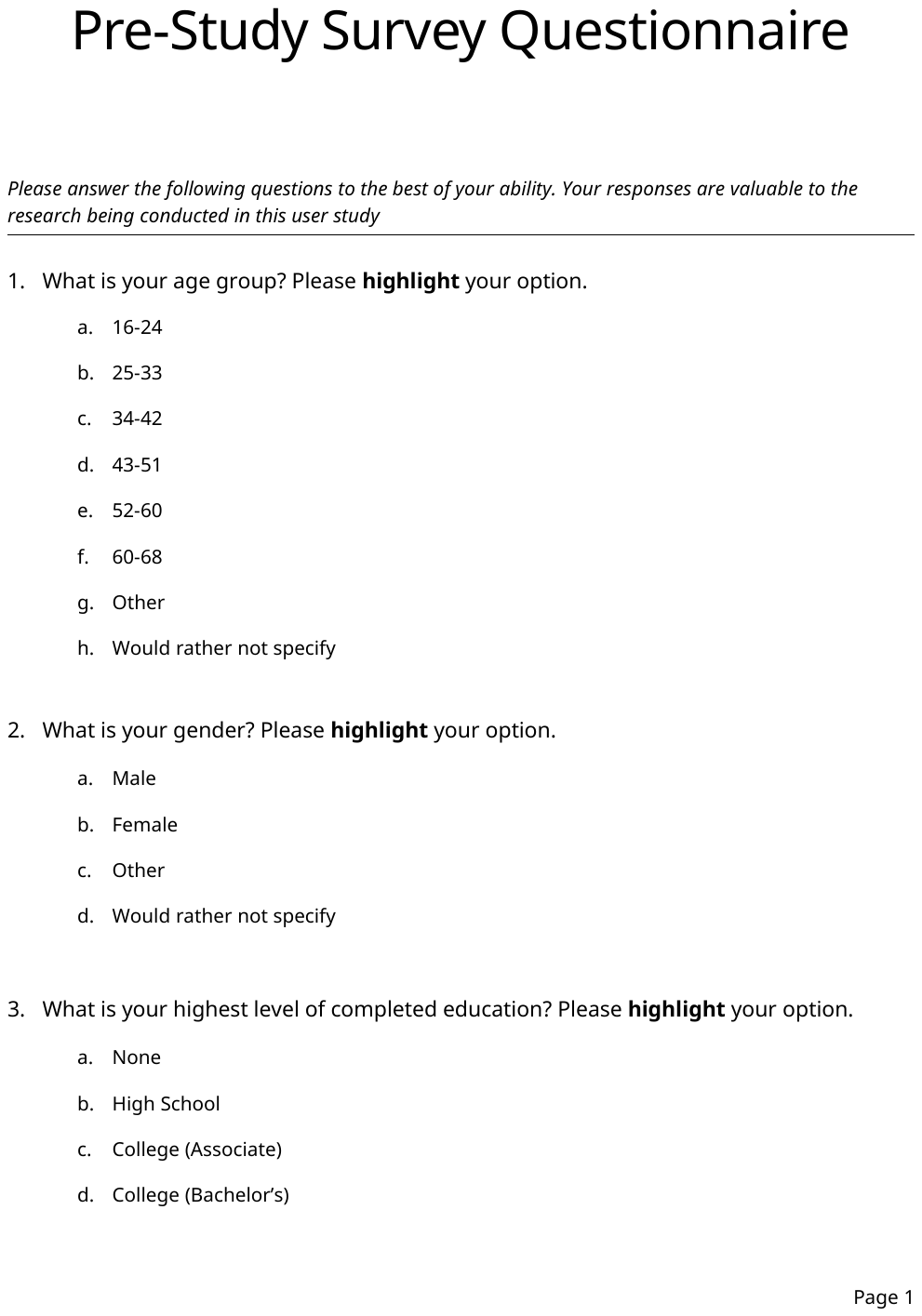}
\section{Post-Study Survey Questionnaire}\label{"apx:post"}

\includepdf[pages=-]{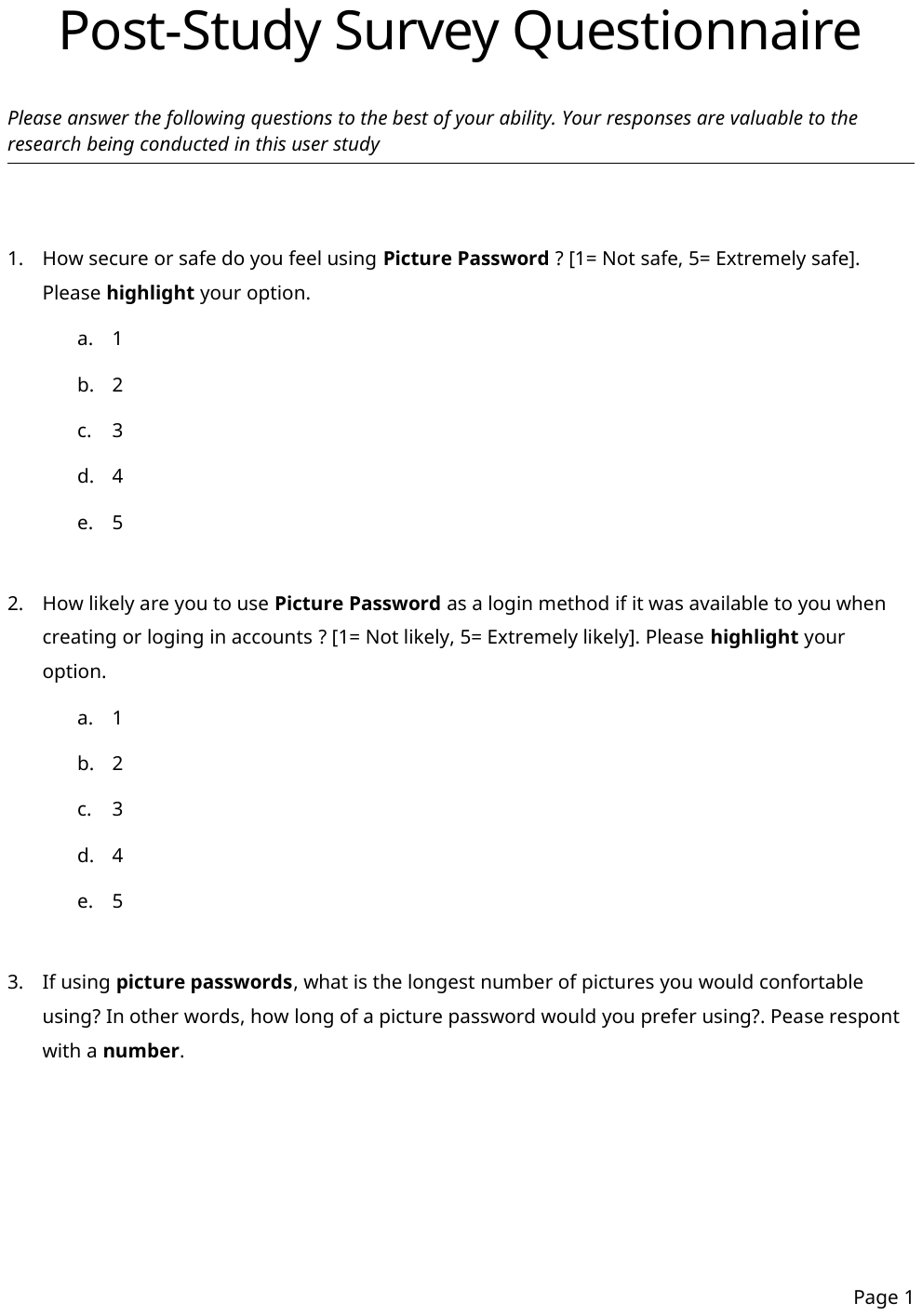}

\end{document}